\documentclass[acmsmall,screen,nonacm]{acmart}
\acmJournal{PACMPL}
\acmVolume{1}
\acmNumber{CONF} % CONF = POPL or ICFP or OOPSLA
\acmArticle{1}
\acmYear{2018}
\acmMonth{1}
\acmDOI{} % \acmDOI{10.1145/nnnnnnn.nnnnnnn}
\startPage{1}

\setcopyright{cc}
\setcctype{by}
\copyrightyear{2023}           %% If different from \acmYear
\usepackage{booktabs}   %% For formal tables:
\usepackage{subcaption} %% For complex figures with subfigures/subcaptions
\usepackage{hyperref}
\usepackage{tabularx}

\newcommand*{\toolname}{REVIS}

\begin{document}

%% Title information
\title[\toolname]{\toolname: An Error Visualization Tool for Rust}         %% [Short Title] is optional;
                                        %% when present, will be used in
                                        %% header instead of Full Title.
%\titlenote{with title note}             %% \titlenote is optional;
                                        %% can be repeated if necessary;
                                        %% contents suppressed with 'anonymous'
%\subtitle{Subtitle}                     %% \subtitle is optional
%\subtitlenote{with subtitle note}       %% \subtitlenote is optional;
                                        %% can be repeated if necessary;
                                        %% contents suppressed with 'anonymous'

%% Author information
%% Contents and number of authors suppressed with 'anonymous'.
%% Each author should be introduced by \author, followed by
%% \authornote (optional), \orcid (optional), \affiliation, and
%% \email.
%% An author may have multiple affiliations and/or emails; repeat the
%% appropriate command.
%% Many elements are not rendered, but should be provided for metadata
%% extraction tools.

%% Author with single affiliation.
\author{Ruochen Wang}
%\authornote{with author1 note}          %% \authornote is optional;
                                        %% can be repeated if necessary
%\orcid{0009-0005-1083-5591}             %% \orcid is optional
\affiliation{
%  \position{Position1}
%  \department{Department1}              %% \department is recommended
  \institution{University of California, San Diego}            %% \institution is required
%  \streetaddress{Street1 Address1}
%  \city{City1}
%  \state{State1}
%  \postcode{Post-Code1}
  \country{USA}                    %% \country is recommended
}
\email{wangrc@ucsd.edu}          %% \email is recommended

%% Author with two affiliations and emails.
\author{Molly MacLaren}
%\authornote{with author2 note}          %% \authornote is optional;
                                        %% can be repeated if necessary
%\orcid{nnnn-nnnn-nnnn-nnnn}             %% \orcid is optional
\affiliation{
%  \position{Position2a}
%  \department{Department2a}             %% \department is recommended
  \institution{University of California, San Diego}           %% \institution is required
%  \streetaddress{Street2a Address2a}
%  \city{City2a}
%  \state{State2a}
%  \postcode{Post-Code2a}
  \country{USA}                   %% \country is recommended
}
\email{mmaclaren@ucsd.edu}         %% \email is recommended

\author{Michael Coblenz}
%\orcid{0000-0002-9369-4069}             %% \orcid is optional
\affiliation{
%  \position{Position2b}
%  \department{Department2b}             %% \department is recommended
  \institution{University of California, San Diego}           %% \institution is required
%  \streetaddress{Street3b Address2b}
%  \city{City2b}
%  \state{State2b}
%  \postcode{Post-Code2b}
  \country{USA}                   %% \country is recommended
}
\email{mcoblenz@ucsd.edu}         %% \email is recommended

%% Abstract
%% Note: \begin{abstract}...\end{abstract} environment must come
%% before \maketitle command
\begin{abstract}
Rust is a programming language that uses a concept of ownership to guarantee memory safety without the use of a garbage collector. However, some error messages related to ownership can be difficult to understand and fix, particularly those that depend on value \emph{lifetimes}.
To help developers fix lifetime-related errors, we developed \toolname, a VSCode extension that visualizes lifetime-related Rust compiler errors.
We describe the design and implementation of the VSCode extension, along with a preliminary evaluation of its efficacy for student learners of Rust. Although the number of participants was too low to enable evaluation of the efficacy of \toolname{}, we gathered data regarding the prevalence and time to fix the compiler errors that the participants encountered.

\end{abstract}

%% 2012 ACM Computing Classification System (CSS) concepts
%% Generate at 'http://dl.acm.org/ccs/ccs.cfm'.

%% End of generated code

%% Keywords
%% comma separated list
  %% \keywords are mandatory in final camera-ready submission

%% \maketitle
%% Note: \maketitle command must come after title commands, author
%% commands, abstract environment, Computing Classification System
%% environment and commands, and keywords command.
\maketitle

\section{Introduction}

Rust is a programming language that enforces memory safety without the need for a garbage collector.
However, using Rust requires that programmers learn challenging concepts pertaining to Rust's type system.
Some error messages emitted by the Rust compiler depend on understanding Rust-specific concepts of \emph{ownership} and \emph{lifetimes}, which may be challenging for novices. These challenges may impede adoption of Rust~\cite{fulton2021}, resulting in users continuing to use unsafe languages such as C and C++.

In this paper, we describe \toolname{} (the Rust Error Visualizer), a new Visual Studio Code extension that provides visualizations for error messages that pertain to a Rust concept called \emph{lifetimes}, which relate to Rust's notion of ownership.
In Rust, every value has a unique owner, which enables the compiler to automatically manage memory allocation for that value.
Each value has a \emph{lifetime}, which starts when the value is allocated and ends when it is no longer used, at which point it can be automatically deallocated.
Because these lifetimes span multiple lines of code, error messages about lifetimes relate to spans of code.
These are different from traditional compiler error messages, which usually pertain to individual lines of code. 

Ownership of a value can be \emph{moved} to a new variable.
When a value is assigned to a new variable or passed as an argument to a function, ownership of that value is transferred from the previous owner to the new owner.
Ownership can also be \emph{borrowed} via references, which can be either immutable or mutable.
While a value is borrowed, the owner retains ownership and the borrower has a limited scope of access.
The borrow checker in the Rust compiler enforces a set of rules on variable lifetimes to avoid invalid references and race conditions. Different instances of the same variable can have different types, since borrow-checking is a flow-sensitive analysis. We argue that custom visualizations may be more appropriate for showing errors that result from flow-sensitive analyses than traditional textual approaches.
%The complexity of the system results in error messages that relate to sequences of operations across multiple lines of code.

\begin{figure}[htb]
    \centering
    % TODO: triangle misaligned after vscode upgrade
    \includegraphics[width=0.9\textwidth]{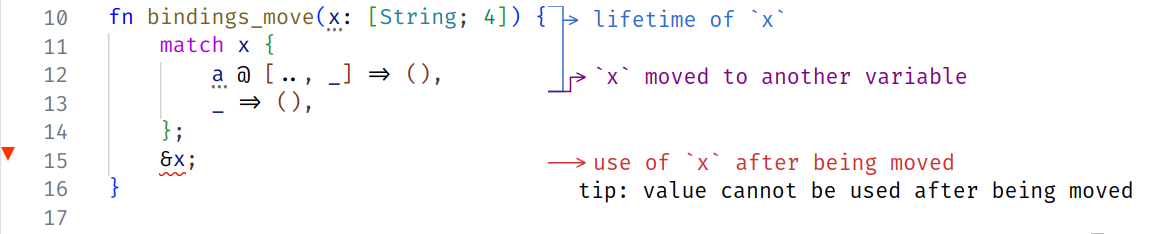}
    \caption{\toolname{}'s visualization for a use after move error. The function argument \texttt{x} is defined in line 10 and moved to variable \texttt{a} in line 12, so its lifetime is lines 10 -- 12. The usage of \texttt{x} on line 15 is an error because it occurs after line 12. The visualization shows that the use of \texttt{x} occurs outside of \texttt{x}'s lifetime.}
    \label{fig:example597}
\end{figure}

Figure~\ref{fig:example597} shows an example of how \toolname{} works.
The red triangle in the gutter indicates there is a visualization available on this line; the triangle points downward to indicate that the visualization is currently displayed.
%The red squiggly underline under \texttt{\&x} is displayed by rust-analyzer.
In the diagram at the right, the blue region shows the lifetime of the variable \texttt{x}. The red arrow shows the use of \texttt{x}, which is erroneous because it is outside the blue region.
The purple arrow shows why the lifetime of \texttt{x} ends: its value was moved to another variable. A short tip is shown below the diagram to explain the cause of the error.

\toolname{} is motivated by the difficulty novices have learning the rules of Rust's type system. Novice Rust programmers often find lifetime-related compile errors difficult to understand, even when the error messages that the Rust compiler produces are generally thought to be excellent~\cite{fulton2021,Crichton2020:Usability}.
Unlike many kinds of errors, lifetime-related errors pertain to spans of code, not just individual lines. Prior programming tools have little special accommodation for errors that include multiple regions of code that relate to each other.
They typically are able to display decorations for ranges of characters with associated messages, but the decorations may not be clearly related to each other, particularly because multiple errors can occur in one region of code.
In addition, the associated messages cannot be displayed together, so developers cannot see the complete information provided by the error messages. \toolname{} uses disclosure triangles in the gutter to allow programmers to focus on all the information pertaining to individual errors.

We conducted a preliminary evaluation of \toolname{} in a Rust-based compilers course at our university. In our evaluation, we randomly assigned participants to use or not use \toolname{}. We collected snapshots of each version of their code that was built, enabling us to analyze error occurrence rates and costs of fixing each error. Although we did not recruit enough participants to enable a quantitative evaluation of \toolname{}'s benefits, we found that four of the twenty most time-intensive errors to fix pertained to borrowing or ownership. This constituted about 3\% of the time spent fixing errors, suggesting that making a significant impact on error-fixing time may require additional kinds of tool or educational support. If this fraction generalizes to other programming contexts, it could mean that fixing Rust-specific compiler errors does not present a significant barrier to Rust adoption. Alternatively, addressing lifetime-related errors may be more important in a broader programming context compared with the restricted domain of a compilers course.

The contributions of this paper can be summarized as follows:
\begin{itemize}
    \item We describe \toolname, a novel error visualization tool for Rust lifetime errors.\footnote{The tool is available in VSCode marketplace at \url{https://marketplace.visualstudio.com/items?itemName=weirane.errorviz}.}
    \item We conducted a preliminary evaluation of \toolname{}, observing that lifetime and borrowing errors occur frequently but only consume about 3\% of the total error-fixing time in the context of the compilers course in which we did the study.
\end{itemize}

\section{\toolname{} Tool}

\subsection{Supported Errors}

The Rust compiler produces many kinds of lifetime-related errors. \toolname{} focuses on those errors that have all the related information within the same function. \toolname{} supports eight errors, including use of variables outside of their lifetimes, variable use or borrowing when already mutably borrowed, and variables not living long enough.

%\mc{I think this is too much detail to include here. Instead, summarize.}
%Currently, our tool supports the following error codes:
%\begin{itemize}
%    \item E0373: A captured variable in a closure may not live long enough.
%    \item E0382: A variable was used after its contents have been moved elsewhere.
%    \item E0499: A variable was borrowed as mutable more than once.
%    \item E0502: A variable already borrowed as immutable was borrowed as mutable.
%    \item E0503: A value was used after it was mutably borrowed.
%    \item E0505: A value was moved out while it was still borrowed.
%    \item E0506: An attempt was made to assign to a borrowed value.
%    \item E0597: A value was dropped while it was still borrowed.
%\end{itemize}

\subsection{Visualization Design}
\label{sec:design}

\begin{figure}[htb]
    \centering
    \includegraphics[width=\textwidth]{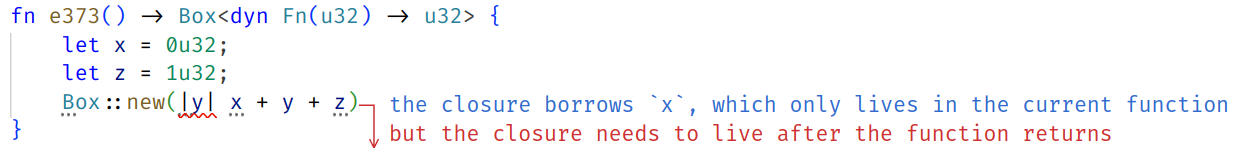}
    \caption{Visualization with an open region. The function returns a closure that captures a local variable. Because the lifetime of the local variable does not cover the lifetime of the closure, there is an error. The region is open at line 15 because it suffices to show that the lifetime of the closure extends beyond the function.}
    \label{fig:oneend}
\end{figure}

\toolname{}'s visualizations consist of three types of components: \emph{regions}, \emph{arrows} and \emph{tips}. Regions are vertical lines that span across multiple lines of code with horizontal end points on each end and descriptive text to the right.
They are mainly used to show the lifetimes of variables.
In figure~\ref{fig:example597}, the blue component that spans from line 10 to line 12 is a region.
If only one end of the region is relevant to the error, the region will be open and have one horizontal end point and an arrow pointing up or down at the other end, as figure~\ref{fig:oneend} shows.

\emph{Arrows} are horizontal lines with arrowheads pointing to the right.
Each arrow is attached to a line of code and indicates a single event, such as a borrow or move, that is relevant to the error.
The head of the arrow is always positioned at the center of the line while the tail can be configured to start at any position.
Because the ends of regions are not at the center of a line, allowing arrow tails to start at any position can make the visualizations easier to read.
For example, in figure~\ref{fig:example597}, the tail of the purple arrow is positioned to overlap with the end of the blue region.

\emph{Tips} are lines of text at the bottom of the visualization.
They explain why the error occurred and may give suggestion on how to fix the error.

%Rust compiler output for lifetime-related errors typically contains descriptions about lifetime regions and variable move or borrow actions.
%Variable lifetime spans across multiple lines and move or borrow is an action that occurs on a single line.
%To reflect the two types of descriptions\mc{what are these?}, our visualizations are composed of two types of components: \emph{regions} and \emph{arrows}.
%Visualization may also contain \emph{tips}.

The components have two severity levels: error and information, matching the severity levels of Rust compiler messages.
Error-level components are displayed in red and information-level components are displayed in blue or purple.

\subsection{Implementation}

Our extension depends on the rust-analyzer extension~\cite{rustanalyzer}. When a source file is saved, rust-analyzer runs the Rust compiler to obtain compile errors for the current file.
This storing action triggers our ``save diagnostics'' hook on diagnostics change.
Inside the hook, we first filter out Rust errors with supported error codes from the diagnostics list, and then display right-pointing red triangles for each of them on the corresponding lines so the user knows there is a visualization available.
The error code, position, and message of the error are stored in this step for use when generating the visualization.

When the cursor is on the line with the triangle, a keyboard shortcut causes our extension to display the visualization and rotates the triangle to indicate that the visualization is shown. Later, to hide the visualization, the user can press the same keyboard shortcut again.
Visualizations are SVG images that consist of the components described in section~\ref{sec:design}.
We compute visualizations by parsing the error message according to the error code.
As errors with the same error code have a consistent error message pattern, for errors with a certain error code, we extract variable lifetime information and move/borrow actions from the message, and use \emph{region} or \emph{arrow} to display them.
After the SVG image is computed, we calculate the maximum width of the lines on which the image will take and make sure the image does not overlap with the code.
The image is configured as a decoration of the first line that the image is displayed.
When the user zoom in or out their editor view, the image is able to scale with the text so the visualization will align with the code.

\toolname{} uses keyboard shortcuts to open and close visualizations because VSCode does not support mouse clicks on the gutter and the visualization image yet~\cite{clickissue}.
If the user saves the file again, displayed visualizations are  automatically updated.

\section{Evaluation}
% 1 method 2 results { 1 quantitative 2 qualitative? objective }
% edit abstract and intro to include preliminary eval
We ran a user study in a combined graduate-undergraduate compiler construction course during spring quarter 2023. In the course, students were introduced to Rust for the first time and completed several programming assignments over the duration of the class.

\subsection{Method}

Students who agreed to participate installed a \texttt{build.rs} file, which committed each built version of their code to a remote repository. It happens when the file is saved and the Rust compiler is run. We collected work for approximately four weeks, starting six weeks into the ten-week term (due to a delay in IRB review). After the study, the participants completed an exit survey regarding their programming experience, and their experience using the tool. Students who contributed code and filled out the survey received a \$5 gift card. In addition, students could opt in to be a randomly assigned to either use \toolname{} or not; students who opted in received an additional \$5 gift card. Our study was approved by our IRB.

\subsection{Results}

We gathered commit information from six students, five of whom had professional software development experience. Five agreed to random assignment; three actually used \toolname{}.

\paragraph{Exit Survey}

Two of the three students assigned to use the tool said it was useful and are likely to use it again, while one was unable to access the visualizations. The participants rated the relative difficulty (on a Likert scale) of \toolname{}-supported errors. E0597, in which a variable with a borrowed value does not live long enough, had four out of seven responses range from "Moderately hard" to "Very hard." Other errors such as E0382, the result of a borrow of a moved variable, were either rated as "Very easy" or were never encountered by five of the seven participants.
%Without a visualization tool, four of the seven participants said that fixing compiler errors in Rust was "a little harder" to "a lot harder" than in Javascript, where the other three said "about the same" or had not used Javascript before. With a visualization tool, the two students who used the visualizations said fixing errors was "a little easier" and "a lot easier" than in Javascript.

\begin{figure}
    \centering
    \includegraphics[width=0.9\textwidth]{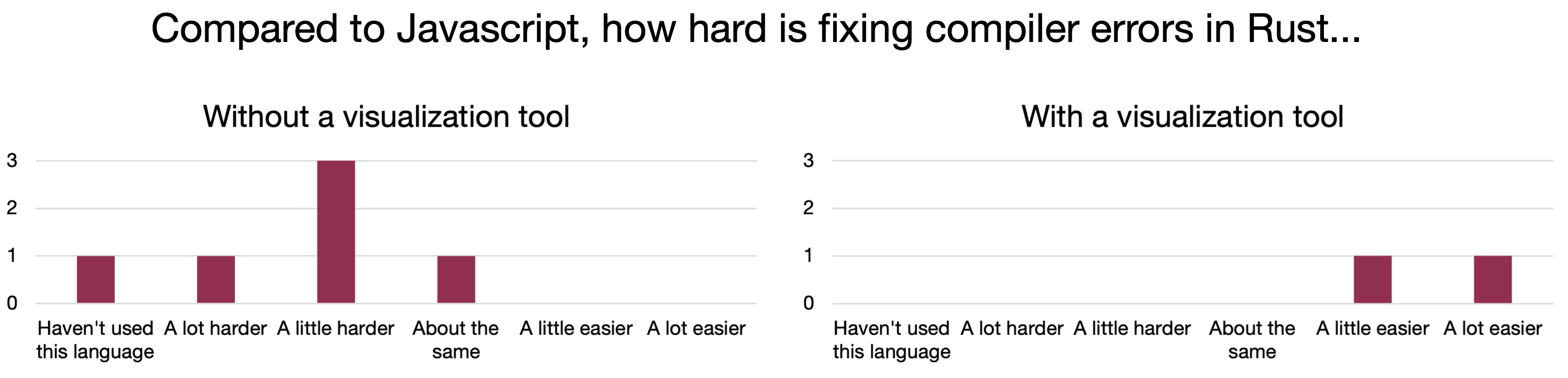}
    \caption{Participants who had access to the visualizations were asked to compare the difficulty of fixing compiler errors with and without \toolname{} to the difficulty in other languages.}
    \label{fig:jschart}
\end{figure}

\paragraph{Quantitative Analysis}
We gathered 6334 commits from six students, three of whom used \toolname{} and three of whom did not. However, only 833 of those commits came from students who used the tool, as some students did not continue to submit their progress throughout the study. We built each version and parsed the error messages and repository timestamps into a SQLite database.

In order to determine time taken to fix distinct compilation errors, we adapted Mesbah et al.'s methodology for determining Active Resolution Cost (ARC)~\cite{mesbah2019}. A \emph{resolution session} is a sequence of consecutive builds, $B_1, B_2, \dots, B_k$ where a given error message $E_i$ first appears in $B_1$ and is resolved in $B_k$. If two consecutive builds are separated by at least 1500 seconds, we assume the student may have taken a break, so we cap the time at 1500 seconds.

We simplified the definition of ARC by \emph{not} excluding build time due to the small scale of the programming assignment. The ARC for error $E_i$ with resolution session $B_1, B_2 \dots, B_k$ is defined in equation \ref{arcdef}.

\begin{equation}
\label{arcdef}
\mathrm{ARC} \ \stackrel{\text{def}}{=} \ \sum_{i=1}^{k-1} \frac{T_i}{|E_i|}
\end{equation}

We categorized the data by error code to count the total number of occurrences, total time spent fixing each error, and average time spent fixing errors by error code across participants. Because individual times are estimated, we refer to them as \emph{costs} in Table \ref{table:costtable}.

\begin{table}[b]
\caption {Top 20 Error Codes by Active Resolution Cost. ``Total" column shows the fraction of the total cost that was due to fixes for this error. Errors that are supported by \toolname{} are shown in bold.}
\begin{tabularx}{\columnwidth}{l X l l l l}
\toprule
\textbf{Error} & \textbf{Description} & \textbf{\#} & \multicolumn{3}{c}{\textbf{Cost}}  \\ 
\cmidrule{4-6}
& & & \textbf{Total}  & \textbf{Avg} & \textbf{Std. Dev.} \\
\midrule
E0308 & Expected type did not match the received type & 360 & 26\%  & 69s & 146s \\
E0425 & An unresolved name was used & 363 & 22\% & 58s & 156s \\
N/A & Syntax error & 275 & 9.7\% & 33s & 71s\\
E0599 & Method is used on a type which doesn't implement it & 145 & 9.2\% & 60s & 153s \\
E0004 & Compiler cannot guarantee a matching pattern & 63 & 7.3\% & 110s & 137s \\
E0061 & Invalid \# of arguments passed to function & 92 & 4.7\% & 48s & 100s \\
E0277 & Use of type that does not implement some trait & 109 & 3.7\% & 32s & 44s\\
E0063 & Struct's or struct-like enum variant's field not provided & 24 & 2.0\% & 79s & 253s \\
E0433 & An undeclared crate, module, or type was used & 67 & 1.4\% & 19s & 64s\\
E0507 & A borrowed value was moved out & 21& 1.1\%  & 49s & 93s \\
E0609 & Attempted to access a non-existent field in a struct & 27 & 0.99\% & 34s & 39s \\
E0412 & A used type name is not in scope & 48 & 0.96\% & 19s & 31s \\
\textbf{E0597} & A value was dropped while it was still borrowed & 7 & 0.88\% & 119s & 51s \\
E0133 & Unsafe code used outside of an unsafe context & 11 & 0.82\%  & 70s & 173s\\
E0596 & You tried to mutably borrow a non-mutable variable & 10 & 0.74\%  & 69s & 121s \\
\textbf{E0382} & Variable used after its contents were moved elsewhere & 16 & 0.66\% & 39s & 27s \\
E0432 & An import was unresolved & 28 & 0.58\% & 20s & 46s \\
E0614 & Dereferenced a variable which cannot be dereferenced & 15 & 0.40\% & 25s & 36s \\
E0282 & Compiler could not infer type; need type annotation & 9 & 0.40\% & 42s & 49s \\
E0369 & Binary operation on a type which doesn't support it & 11 & 0.38\% & 32s & 42s\\
\bottomrule
\end{tabularx}
\label{table:costtable}
\end{table}

\begin{figure}
    \centering
    \includegraphics[width=0.9\textwidth]{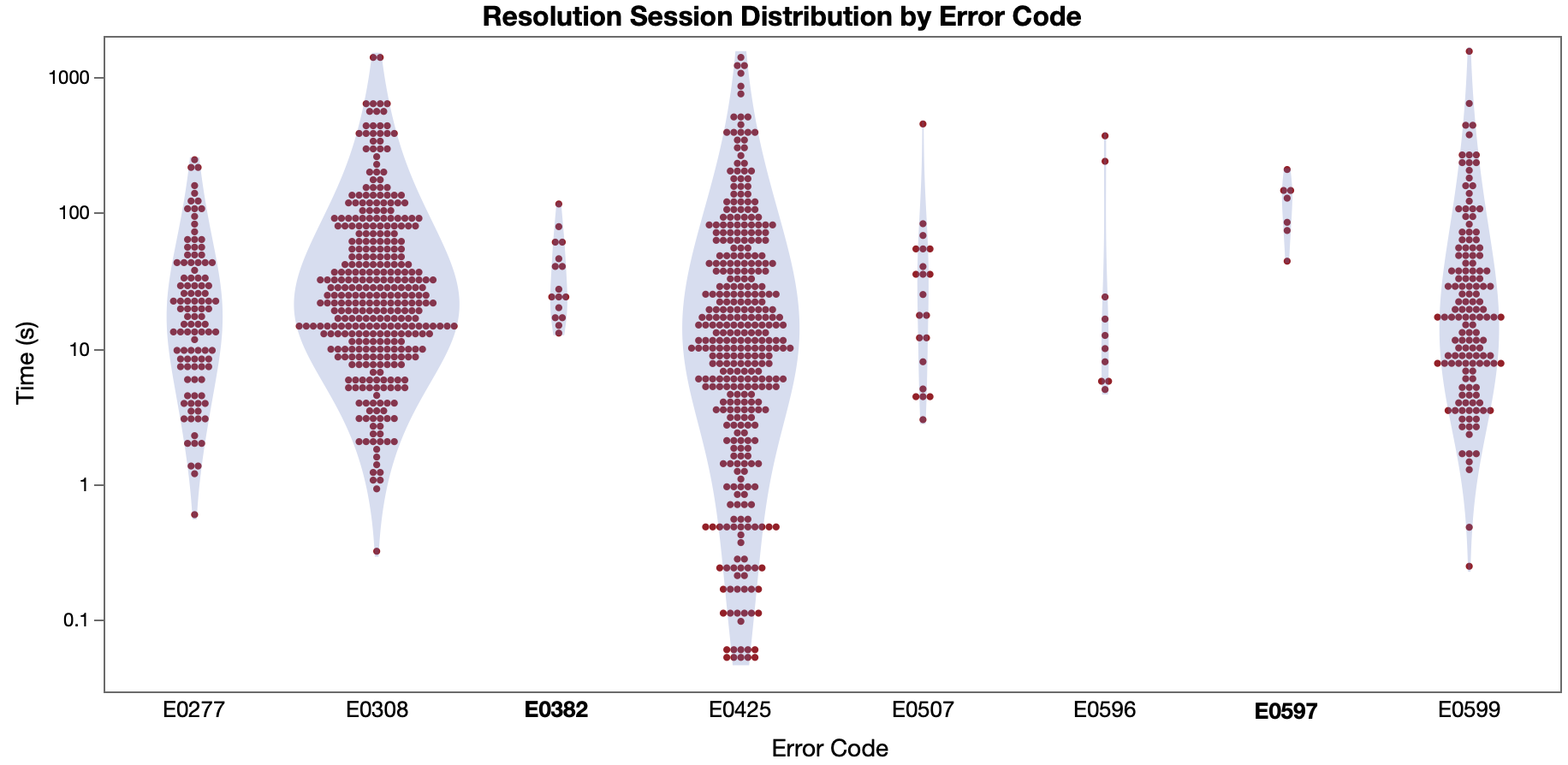}
    \caption{Log-scale violin plot representing the time taken to resolve errors by error code. Each dot represents a resolution session. Both E0382 and E0597 have a lower bound > 10 seconds.}
    \label{fig:vplot}
\end{figure}

Of the 15019 total error messages, 51 were supported by \toolname{}, and of the 51, only 5 came from students using the tool with 3 distinct resolution sessions. In the context of this compilers assignment, borrowing and ownership-related errors did not appear as frequently as other errors such as E0425: \texttt{An unresolved name was used}, which has the most resolution sessions.
 %The most frequent errors being those that relate to the compiler's inability to resolve a particular variable
 Unresolved variable name errors being the most frequent compiler error is consistent with Mesbah's results with Java as well as \cite{seo2014}'s results with C++.

Table~\ref{table:costtable} shows the number of resolution sessions as well as the total, average, and standard deviation of calculated Active Resolution Cost by error code. In terms of total cost, E0597 ranked at \#13 and E0382 at \#16 both can be visualized using \toolname{}. In addition, E0597 has the \#4 highest average time. These two errors make up only 1.5\% of the total cost. However, combined with other ownership-related errors not yet supported by \toolname{} like E0507 and E0596, they contribute to about 3.4\% of repair time. Other errors such as those involving traits are also unique to Rust. E0227 consumed 3.7\% of the total time. Together, these five errors contribute to about 7.1\% of the total cost.

\section{Limitations}

The preliminary evaluation took place in a compilers course, in which the work may not reflect general-purpose programming work,
because the compiler was written in a functional style and most of the references were immutable references.
The study began about five weeks into the course, at which point the participants (who were students) already had a certain amount of expertise in Rust. These two factors, combined with the limited set of participants, significantly limit the external validity of the study. Further work is needed to evaluate \toolname{} in a more general context.

\section{Discussion}

Participants spent only a limited amount of time fixing lifetime-related errors (3.4\% of the total error-fixing time). However, we suspect that by the time the study ran, the development challenges that would have resulted in most lifetime-related errors were already resolved, since the overall structure of the compiler was already set. Alternatively, it is possible that the challenges of Rust's type system have an overstated significance in the prior literature.

In spite of the relatively small amount of time spent fixing lifetime-related errors, borrow errors supported by \toolname{} appear to take longer to fix than other errors. Tools to make fixing these errors easier may help reduce user frustration and the perception of type system challenges even if they do not make a significant impact on task completion times.

One borrowing error, \emph{value dropped while borrowed}, may be much harder to fix than other borrowing errors. This may represent an opportunity for future tool development.

\section{Future Work}

We plan to conduct a larger-scale evaluation of our tool that focuses on a more general setting rather than a classroom setting. 

We hope to improve \toolname{} to support more error codes related to lifetimes, including errors that are not constrained in single functions.
We will need to find a way to relate different parts of the code that are potentially far away from each other.
It will also be useful to enrich the messages in the visualizations, as we only use the existing output of the Rust compiler to compute the visualizations.
For example, in figure~\ref{fig:example597}, we could show the users ``\texttt{\textasciigrave x\textasciigrave{} moved to \textbf{\textasciigrave a\textasciigrave}}'' instead of ``\texttt{\textasciigrave x\textasciigrave{} moved to another variable}.''
We also want to include an opt-in data collection component in our tool and release it to the public to allow us to conduct a study in the real world.
The data collection component will also be able to give users insight of their coding behavior.

E0507 (borrowed value was moved out) and E0596 (tried to mutably borrow a non-mutable variable) may represent opportunities for future work; although these are not lifetime-related errors, they are specific to sequences of ownership operations. A future extension may be able to propose fixes, helping users address these errors more quickly. Alternatively, better educational tools could help users avoid introducing these errors.

% \todo{ help people identify when the really need to learn concepts, vs. when concepts are mentioned due to a more general case}

Since one participant had trouble seeing the visualizations, we may consider opening them by default. In the future we hope VSCode will enable detection of clicks in the gutter.

%As an extension of the tool, we want to create a more comprehensive Rust programming helper

%We can also explore the use of visualization on error messages of other programming languages.

\section{Related Work}

Several authors have described empirically-evaluated guidelines for writing textual error messages. Denning et al.~\cite{Denny2021:Designing} found that length, jargon usage, sentence structure, and vocabulary are affect textual error message readability. Becker et al.~\cite{Becker2021:Towards} found in an experiment that error message readability can be subjective and experience-dependent, but generally participants believed that length, tone, and jargon usage affected readability. Becker et al.~\cite{Becker2021:Towards} summarized many earlier results, among with is the argument that good error messages can make a large impact on usability for beginners. Our design of \toolname{} particularly reflects their recommendation \emph{provide context to the error}, since it aligns its visualizations with the relevant erroneous code. It also reflects the recommendation \emph{increase readability}, since the diagrams reduce the amount of additional text needed relative to text-only errors. 

%\cite{becker2021} shows that experts, non-experts and students assess readability of error messages differently.

\begin{figure}[htp]
    \centering
    \includegraphics[width=0.9\textwidth]{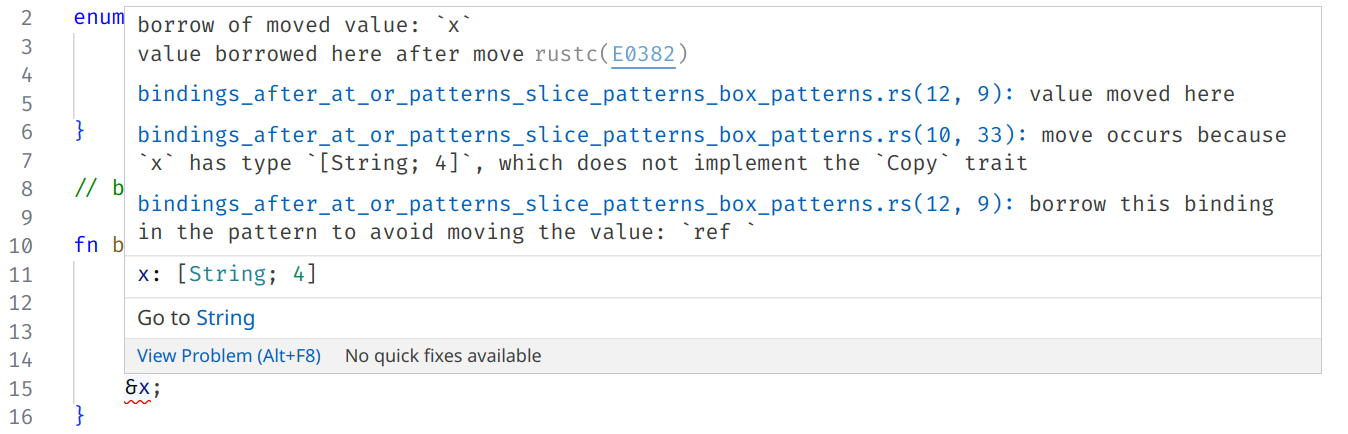}
    \caption{Rust-analyzer hover box displayed for the erroneous ``\texttt{\&x}'' in figure~\ref{fig:example597}}
    \label{fig:withoutviz}
\end{figure}

Existing editor tools around Rust errors have little special support for errors that span multiple lines.
Rust-analyzer~\cite{rustanalyzer} is a popular Rust language extension for VSCode. As shown in figure~\ref{fig:withoutviz}, the rust-analyzer VSCode extension displays multiple decorations for errors: red squiggly underline for the erroneous code and gray dotted underlines for code related to the errors.
When the developer moves the mouse onto one of the decorations, the hover box that pops out contains a textual description of all the related information as line numbers and text.
It cannot display all the hover boxes related to one error together, which is a problem because the various contributing lines can have their own boxes. 
Also, the hover boxes may contain irrelevant information, such as the type of the variable under the cursor.
Finally, they may overlap with the code that caused the error. \toolname{} always puts its diagrams to the right of the code, avoiding any overlap.

\begin{figure}[htp]
    \centering
    \includegraphics[width=0.9\textwidth]{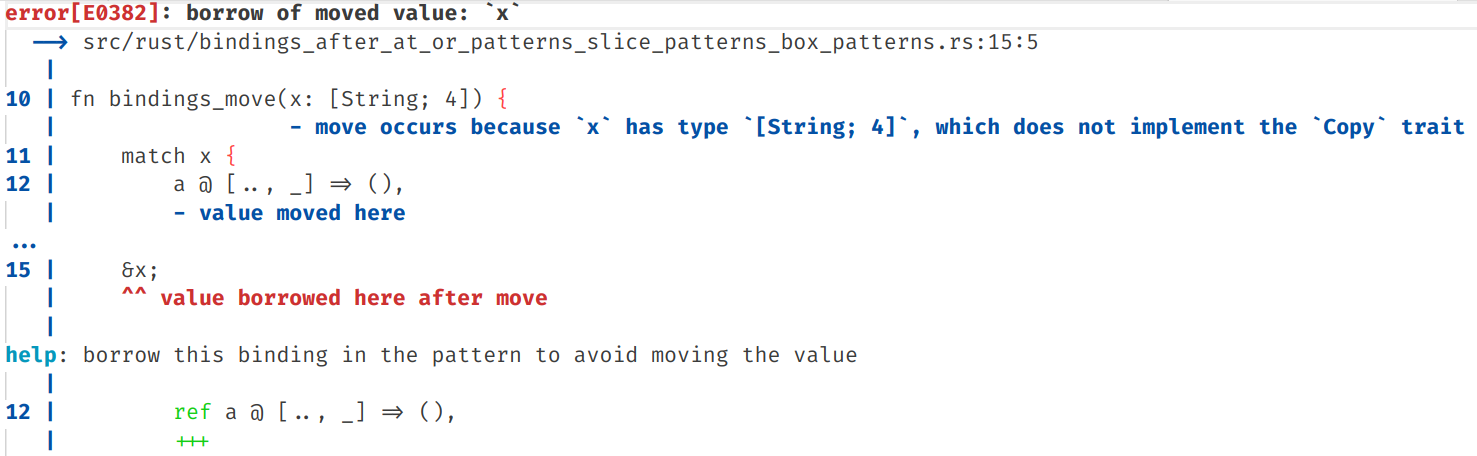}
    \caption{Command line error message for the error in figure~\ref{fig:example597}}
    \label{fig:clierror}
\end{figure}

Developers also view command line error messages directly in the terminal.
Command line error messages are able to convey all the diagnostics in one place, as shown in figure~\ref{fig:clierror}.
However, because the descriptions are in a separate window than their editor, users need to move back and forth to relate the error messages to the code in their editor.
Sometimes command line output may omit lines of code (lines 13 and 14 are omitted in figure~\ref{fig:clierror}), which makes it more difficult to build the relation between the error on the command line and the code in the editor.
Also, although command line errors use different colors to indicate related error messages, they does not point out the lifetime ranges of variables.
With \toolname{}, not only can users view the description of their errors next to their code, they also get an intuitive sense of the lifetime range, which helps them understand the error.

\begin{figure}[htp]
    \centering
    \includegraphics[width=0.9\textwidth]{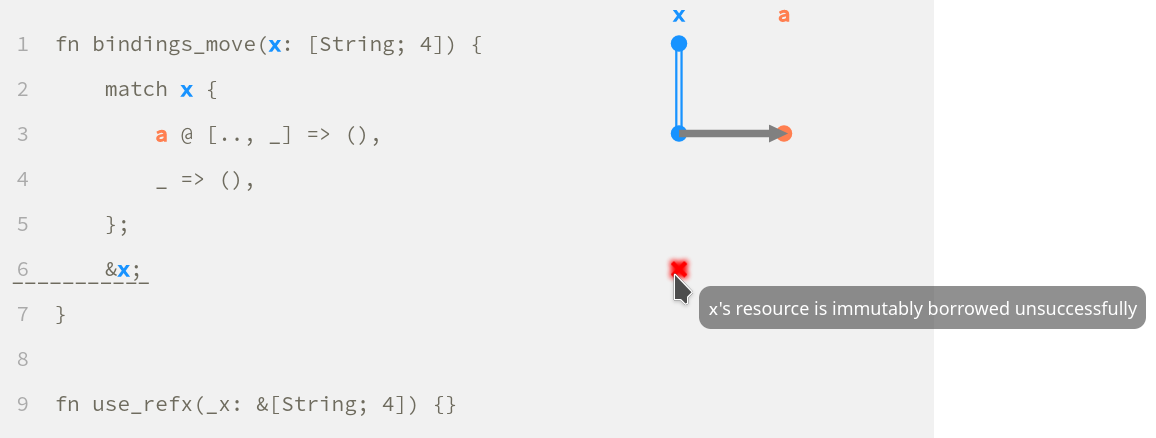}
    \caption{RustViz visualization for the error in figure~\ref{fig:example597}.
    The left column of the visualization shows events of \texttt{x}: it is defined on line 1, moved on line 3 and referenced on line 6.
    The blue vertical line shows the lifetime of \texttt{x} and the gray arrow shows the move from \texttt{x} to \texttt{a}.
    The ``X'' on line 6 shows there is an error with the reference, but the hover information on the ``X'' does not explain why it is an error and how to fix it.}
    \label{fig:rvexample}
\end{figure}

There are existing visualization frameworks for Rust, such as RustViz~\cite{rustviz}.
RustViz enables instructors to create diagrams for Rust programs to help students understand lifetime-related concepts and rules. RustViz could also be used to show errors. For example, figure~\ref{fig:rvexample} shows the same error as in figure~\ref{fig:example597}.
Compared to \toolname{}, RustViz supports a richer set of visual vocabulary, since it supports creating diagrams for arbitrary sequences of Rust ownership events. \toolname{} uses a more limited vocabulary with the goal of making it easier to read the specific errors we wanted it to display. Our approach allows us to explain problems in a more highly-customized way for each error message.

% Hundhausen et al.~\cite{hundausen2017} proposed an IDE-based data analytics process model to help students learn using IDEs.
% The model has four steps: collect data, analyze data, design intervention and deliver intervention.
% The data collected include a wide range of user behavior and output, such as editing/debugging behavior, compiler errors, etc.
% Intervention is the feedback given by the IDE that helps students learn.
% It can come with the form of visualization, notification or constraints.
% In our tool, we collect and analyze the compilation data and produce a visual image to help users understand the error.
% They also suggested that visualizations should present information relevant to learners' immediate tasks, which is why we focus on presenting the error itself.

% \cite{xinogalos2015} investigates what properties introductory programming environment should have.
% They suggested that visualizations of classes, objects and statements are important for students.
% Classes and objects are important language features that are critical for understanding how the language works.
% As an editor extension for introductory Rust programmers, \toolname{} visualizes lifetimes and move/borrow actions, which are also important language features.

%related paper/tools?
%- understanding rust errors
%- visualizing lifetime

%brett becker papers on error messages https://www.brettbecker.com/publications/
\section{Conclusion}
This paper described \toolname{}, a Visual Studio Code plugin that provides visualizations for lifetime-related error messages emitted by the Rust compiler. We conducted a preliminary evaluation in which we collected snapshots of students' work in a compilers course. Although we did not have enough participants to assess whether \toolname{} helped reduce error-fixing time, we observed that most error-fixing time was spent fixing errors that occur in other languages as well. If our data are representative of general Rust development, this could mean that Rust's type system (in terms of time fixing errors) does not impose significant burden compared to other languages.

%% Acknowledgments
\begin{acks}                            %% acks environment is optional
We thank our study participants for submitting their data.
We also thank Roland Rodriguez and Esteban Kuber from Amazon for insights on the future development of \toolname.                                        
\end{acks}

%% Bibliography
\bibliography{main}

%% Appendix
%\appendix
%\section{Appendix}
%
%Text of appendix \ldots

\end{document}